\documentclass[a4paper,11pt]{article}

\usepackage{url}

\usepackage{fullpage}

\usepackage{amsmath,amsthm,amssymb}
\usepackage{xcolor}
\usepackage{graphicx}
\usepackage{bbold}
\usepackage{textcomp} 
\usepackage{enumerate} 
\usepackage{multirow}
\usepackage{overpic}

\DeclareMathOperator{\DIV}{div}
\DeclareMathOperator{\ROT}{rot}

\usepackage{authblk}

\usepackage{todonotes}
\makeatletter
\renewcommand{\todo}[2][]{\tikzexternaldisable\@todo[#1]{#2}\tikzexternalenable}
\renewcommand{\missingfigure}[2][]{\tikzexternaldisable\@missingfigure[#1]{#2}\tikzexternalenable}
\makeatother

\title{Macroscopic Simulation of Isotropic Permanent Magnets}

\author[1]{Florian Bruckner \thanks{florian.bruckner@tuwien.ac.at}}
\author[1]{Claas Abert}
\author[2]{Christoph Vogler}
\author[3]{Frank Heinrichs}
\author[3]{Armin Satz}
\author[3]{Udo Ausserlechner}
\author[3]{Gernot Binder}
\author[3]{Helmut Koeck}
\author[1]{Dieter Suess}
\affil[1]{Christian Doppler Laboratory of Advanced Magnetic Sensing and Materials, Institute of Solid State Physics, Vienna University of Technology, Austria}
\affil[2]{Institute of Solid State Physics, Vienna University of Technology, Austria}
\affil[3]{Infineon Technologies Austria AG, Siemensstrasse 2, 9500 Villach, Austria}

\begin{document}
\maketitle

\begin{abstract}
Accurate simulations of isotropic permanent magnets require to take the magnetization process into account and consider the anisotropic, nonlinear, and hysteretic material behaviour near the saturation configuration. An efficient method for the solution of the magnetostatic Maxwell equations including the description of isotropic permanent magnets is presented. The algorithm can easily be implemented on top of existing finite element methods and does not require a full characterization of the hysteresis of the magnetic material. Strayfield measurements of an isotropic permanent magnet and simulation results are in good agreement and highlight the importance of a proper description of the isotropic material.


\end{abstract}
  
\newpage
\section{Introduction}
Isotropic permanent magnets are used in a wide range of applications, like electric motors and actuators, speakers, as well as magnetic sensors \cite{strnat_modern_1990}. Especially their lower production costs, and the additional capabilities due to the magnetizability in any direction, make them an interesting alternative to high-end anisotropic magnets. Due to the missing preferred direction the magnetization process has a strong influence on the final remanence magnetization within isotropic magnets. Accurate simulations of magnetic components therefore need to consider the non-linear, anisotropic, and hysteretic material behaviour at least close to the perfectly aligned saturation state.

The most famous models for the macroscopic description of magnetic hysteresis are the Preisach model \cite{mayergoyz_mathematical_1986} as well as the Jiles-Atherton model \cite{jiles_theory_1984}. Extensions and combinations of those models have been proposed, where the applicability to the vector hysteresis for isotropic magnets has been demonstrated \cite{mayergoyz_isotropic_1987, koh_vector_2000, daquino_new_2003, fallah_new_2008}. The integration of those material descriptions into finite element simulations is based on either Fixed-Point iteration or the Newton-Raphson method for the solution of the resulting nonlinear systems of equations \cite{saitz_newton-raphson_1999, kuczmann_using_2010, sadowski_inverse_2002}.

Although those methods are well suited for the simulation of dynamic hysteresis effects, the complete description of the hysteresis involves a large computational effort and is not always needed. Considering permanent magnetic materials, the correct description of the hysteretic behaviour near the saturation configuration is often sufficient. In addition to the higher computational effort, the complete characterization of a magnetic material requires complicated reversal curve measurements, which are not always available for commercial magnets. 

Within this work a new method for the description of hysteretic behaviour of isotropic permanent magnets is introduced. The presented algorithm allows to describe small deviations from the saturation configuration in a phenomenological way by means of a single parameter - the {\it freezing field}. The method is very efficient and easy to implement on top of existing magnetic finite element simulations. Moreover there is no need for complete reversal curve measurements of the magnetic material.

The structure of the paper is as follows: Section \ref{ch:anisotropic_magnets} summarizes the numerical method used to simulate the behaviour of anisotropic magnetic components. Within Section \ref{ch:isotropic_magnets} a simple phenomenological algorithm is introduced which allows to extend the method to isotropic magnets. Finally simulations of the magnetic strayfield of an isotropic permanent magnet are compared with measurement data within Section \ref{ch:measurements}.

\section{Anisotropic Magnets} \label{ch:anisotropic_magnets}
Anisotropic magnets are produced with a magnetic field applied during the production process. The magnetic grains orient into the direction of the applied field. After the sintering or bonding process the applied field is turned off, but the prescribed orientation of the magnetic grains remains. Thus, the crystallites are fixed and even very high applied fields do not affect their orientation. There exists a global preferred direction and it is possible to define a homogeneous intrinsic anisotropic material law for the whole magnet. Assuming a uniaxial anisotropy requires the distinction of longitudinal and transversal behaviour. 

For sake of simplicity the global material law will be linearized around the remanence vector $\mathbf{B}_\text{r}$. Assuming the longitudinal easy axis is oriented into the $z$ axis, results in the following constitutive law
\begin{align} \label{eqn:mat}
\begin{split}
&\mathbf{B}(\mathbf{H}) = \mathbf{B}_\text{r} + \mathbf{\mu} \, \mathbf{H} \\
&\mathbf{B}_\text{r} = \begin{pmatrix} 0 \\ 0 \\ B_\text{r} \end{pmatrix} \quad \mathbf{\mu} = \begin{pmatrix} \mu_\bot & 0 & 0 \\ 0 & \mu_\bot & 0 \\ 0 & 0 & \mu_\parallel \end{pmatrix}
\end{split}
\end{align}, 
with the magnetic flux density $\mathbf{B}$, the total field $\mathbf{H}$, the remanence vector $\mathbf{B}_\text{r}$ and the permeability tensor $\mathbf{\mu}$. Nonlinear behaviour can be described by updating the linearized $\mathbf{\mu}$ depending on the actual local field $\mathbf{H}$. For typical applications the demagnetization field shifts the local operating point in a region of the $\mathbf{B}(\mathbf{H})$-curve, which can be well described with a linear model.

It is important to distinguish between the $\mathbf{B}_\text{r}$-vector, the magnetic flux-density without any total magnetic field, as well as the magnetic remanence $\mathbf{B}(\mathbf{H}_\text{ext} = 0)$, which requires a vanishing external field but includes the magnetic strayfield. Measurements typically yield the magnetic remanence, unless a closed loop measurement is performed, which eliminates the magnetic strayfield and therefor allows direct measurement of the $\mathbf{B}_\text{r}$-vector. The terms "$\mathbf{B}_\text{r}$-vector" as well as "magnetic remanence" are used in the following to distinguish these two quantities. 

The actual magnetization state of the permanent magnet can be calculated by means of the homogeneous, magnetostatic Maxwell equations. These are solved efficiently by FEMME Maxwell, which uses a modified version of the method presented in \cite{bruckner_3d_2012}. The (maybe inhomogeneous) magnetic material is contained within the magnetic domain $\Omega$ and is treated by a finite element (FEM) approach. The homogeneous outer region is handled by a boundary element (BEM) approach which is able to deal with open-boundary problems very efficiently. Maxwell's equations as well as the jump conditions at the boundary of the magnetic parts are given by 
\begin{subequations}
  \begin{align}
    \mathbf{\ROT H} &= \mathbf{j}      \quad \text{in} \; \mathbb{R}^3    \\
    \DIV \mathbf{B} &= 0               \quad \text{in} \; \mathbb{R}^3    \\
    [\![ \mathbf{B} ]\!] \cdot  \mathbf{n} &= 0 \quad \text{on} \; \partial \Omega \\
    [\![ \mathbf{H} ]\!] \times \mathbf{n} &= 0 \quad \text{on} \; \partial \Omega
  \end{align}
\end{subequations}, 
where $[\![ . ]\!]$ denotes the jump of a quantity over the boundary $\partial \Omega$ of the magnetic domain, and $\mathbf{n}$ is the outward pointing surface unit normal. Introducing a reduced scalar potential $u$ and expressing the magnetic field created by the current density $j$ as $\mathbf{H}_\text{ext}$ finally leads to the following system of equations
\begin{align} \label{eqn:H}
  \mathbf{H} = \mathbf{H_{ext}} - \nabla u
\end{align}
\begin{subequations} \label{eqn:u}
  \begin{align}
    \mathbf{\nabla} \cdot \left(\mathbf{\mu} \, \mathbf{\nabla} u \right) &= \mathbf{\nabla} \cdot \left( \mu \, \mathbf{H_{ext}} + \mathbf{B}_\text{r}\right)  & & \text{in} \; \Omega \label{eqn:u+} \\
    \Delta u &= 0  & & \text{in} \; \mathbb{R}^3 \setminus \Omega \label{eqn:u-} \\
    [\![ u ]\!] &= 0  & & \text{on} \; \partial \Omega \label{eqn:jump_u} \\
    [\![\nabla u \cdot \mathbf{\mu} \cdot \mathbf{n} ]\!] &= \left( \mathbf{H_{ext}} \cdot [\![ \mu ]\!] + \mathbf{B}_\text{r} \right) \cdot \mathbf{n}  & & \text{on} \; \partial \Omega \label{eqn:jump_phi}
  \end{align}
\end{subequations}

\section{Isotropic Magnets} \label{ch:isotropic_magnets}
In contrast to anisotropic magnets described in the previous chapter, isotropic magnets are produced without an applied field. The magnetic grains are oriented randomly inside of the magnet, and thus the material has no preferred direction. An advantage of this type of magnet is that it is easier to produce and therefore less expensive. Furthermore those magnets can be magnetized into any desired direction due to their isotropic behaviour. On the other hand the magnetic remanence, the coercive field as well as the energy product of isotropic magnets are significantly reduced compared to anisotropic magnets of the same material. 

One additional drawback of isotropic magnets, which should be emphasized in this chapter, is that there exists no intrinsic anisotropy axis of the material. Instead the anisotropic behaviour of the magnet is created by the application of a magnetic field pulse after the production process. For large enough fields the average magnetic moment within the magnet will point into the direction of the spatially varying total field.
As a consequence the local $\mathbf{B}_\text{r}$-vector within the magnet is not only modified at the peak of the magnetization pulse, where the total field is almost equal to the applied field, but also during the decay of the applied field pulse. The demagnetization field gets more and more influence and the $\mathbf{B}_\text{r}$-vector state gets increasingly inhomogeneous during the reduction of the applied field. When the total field at a particular location comes below a certain threshold, the $\mathbf{B}_\text{r}$-vector freezes. 

This effect leads to a shape dependent $\mathbf{B}_\text{r}$-vector state and therefore requires taking into account the specific magnetization process. This can be realized by introducing a critical freezing field $H_\text{freeze}$ and modify the $\mathbf{B}_\text{r}$-vector locally only if the {\it freezing condition} $\vert \mathbf{H} \vert > H_\text{freeze}$ is fulfilled.
\begin{align} \label{eqn:freezing_condition}
  \mathbf{B}_\text{r} = \left\{ \begin{array}{c l} \vert \mathbf{B}_\text{r} \vert \, \mathbf{\hat{H}} & \quad \text{if} \; \vert \mathbf{H} \vert > H_\text{freeze} \\ \mathbf{B}_\text{r} & \quad \text{otherwise} \end{array} \right.
\end{align}

Assuming a simple Stoner-Wohlfarth model for the individual magnetic grains indicates a freezing field between $H_\text{c} / 2 < H_\text{freeze} < H_\text{c}$, where $H_\text{c}$ is the is the coercive field of the individual particle. Starting from a perfectly aligned $\mathbf{B}_\text{r}$-vector state the applied field is gradually reduced and the total field is calculated from equations \eqref{eqn:H} and \eqref{eqn:u}. The field steps need to be chosen small enough to create self-consistent results. In the following chapter simulation results using the presented procedure are compared with measurement data.

\section{Comparison with Measurements} \label{ch:measurements}
The importance of proper simulation of the magnetization process should be demonstrated by a strayfield measurement of an isotropic magnet as it is used as bias magnet in magnetic sensor applications. The dimensions of the magnet used for the measurement is approximately $7.0$\,mm\,$\times 4.5$\,mm\,$\times 5.5$\,mm. The commercially available plastic bonded magnet consists of a NdFeB-composit material specified in Tbl. \ref{tbl:NdFeB} and is directly moulded onto an integrated sensor \cite{elian_integration_2014}.

\begin{table}[h!]
\centering
\small
\begin{tabular}{|c||c|c|c|c|c|c|c|}
\hline
Rare earth magnets & Energy & \multicolumn{2}{|c|}{\multirow{2}{*}{Remanence}} & \multirow{2}{*}{Coercivity} & Relative & Curie \\
(injection moulded) & product & \multicolumn{2}{|c|}{} & & permeability & Temperature \\
  & $(B*H)_\text{max}$ & $B_\text{r}$ & $H_\text{c}^B$ & $H_\text{c}^M$ & $\mu_\parallel / \mu_0$ & $T_c$ \\
  & [kJ/m$^3$] & [mT] & [kA/m] & [kA/m] & $[1]$ & [$^\circ$C] \\
\hline
\hline
isotropic NdFeB & 43.0 & 505 & 340 & 800 & 1.15 & 305 \\
\hline
\end{tabular}
\caption{\small Summary of the magnetic properties of the plastic bonded rare earth magnet material used for the comparison of measurements and simulations.}
\label{tbl:NdFeB}
\end{table}

\begin{figure}[h!]
  \centering
  \includegraphics[width=0.9\columnwidth]{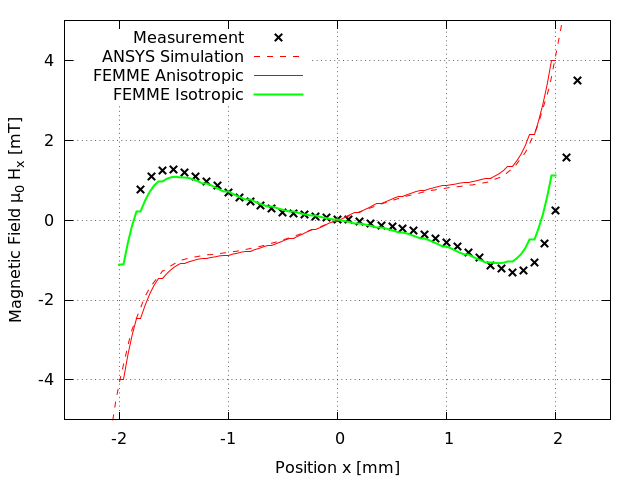}
  \caption{\small Comparison of simulation results with strayfield measurements. Different descriptions of the isotropic permanent magnet are compared: (a) FEMME as well as with (b) ANSYS assuming an anisotropic material law. (c) FEMME using the presented remanence-freezing algorithm for isotropic magnets with a freezing field $\mu_0 H_\text{freeze} = 0.6$\,T.}
  \label{fig:ibb_freeze}
\end{figure}

For the measurement the magnet is magnetized using an impulse magnetizer, which creates a peak magnetic field of $\mu_0 H_\text{max} = 4$\,T. After magnetization in $z$ direction the magnetic strayfield along the $x$-axis is measured $1.0$\,mm above the magnet (see Fig. \ref{fig:ibb_freeze}) using a Hall probe. 

Three different simulations are performed and compared with the measurement results:
\begin{enumerate}[(a)]
  \item \label{itm:femme_anisotropic} {\bf Anisotropic magnet model using FEMME Maxwell \cite{bruckner_3d_2012}} \\
    A homogeneous remanence vector $\mathbf{B}_\text{r} = 0.51$\,T is assumed within the whole magnetic region. A homogeneous, isotropic $\mu = \mu_\bot = \mu_\parallel = 1.12 \, \mu_0$ is chosen since only the longitudinal $\mu_\parallel$ is known from the magnet datasheet. Additional simulations using an anisotropic $\mathbf{\mu}$, with $\mu_\bot = 1.19\,\mu_0$ deduced from measurements, show that the effect of the anisotropic permeability is negligible compared to the shown remanence-freezing effect, which justifies the choice of an isotropic permeability. 
    The magnetization state of the magnet is calculated without external field by means of equations \eqref{eqn:u} combined with the material law \eqref{eqn:mat}.
  \item \label{itm:ansys_anisotropic} {\bf Anisotropic magnet model using ANSYS Maxwell 3D \cite{_ansys_????}} \\
    The commercial and well known electromagnetic field simulation software ANSYS is used to validate the results of the previous simulation. It allows to define anisotropic, hard-magnetic materials using the same parameters $\mathbf{M}_r$ and $\mathbf{\mu}$ as above.
  \item \label{itm:femme_isotropic} {\bf Isotropic magnet model using FEMME Maxwell with remanence-freezing} \\
    For the simulation of the isotropic model it is necessary to incorporate the magnetization process. Therefore an external field of $\mu_0 H_\text{ext} = \mu_0 H_\text{max} = 4$\,T is applied and successively reduced in $0.1$\,T steps until zero external field is reached. Within each step an anisotropic model with the same parameters as above is solve. After each field-step the {\it freezing condition} \eqref{eqn:freezing_condition} is checked and the local $\mathbf{B}_\text{r}$-vector is updated for each element. 
\end{enumerate}

\begin{figure}[h!]
  \centering
  \begin{overpic}[width=0.6\columnwidth]{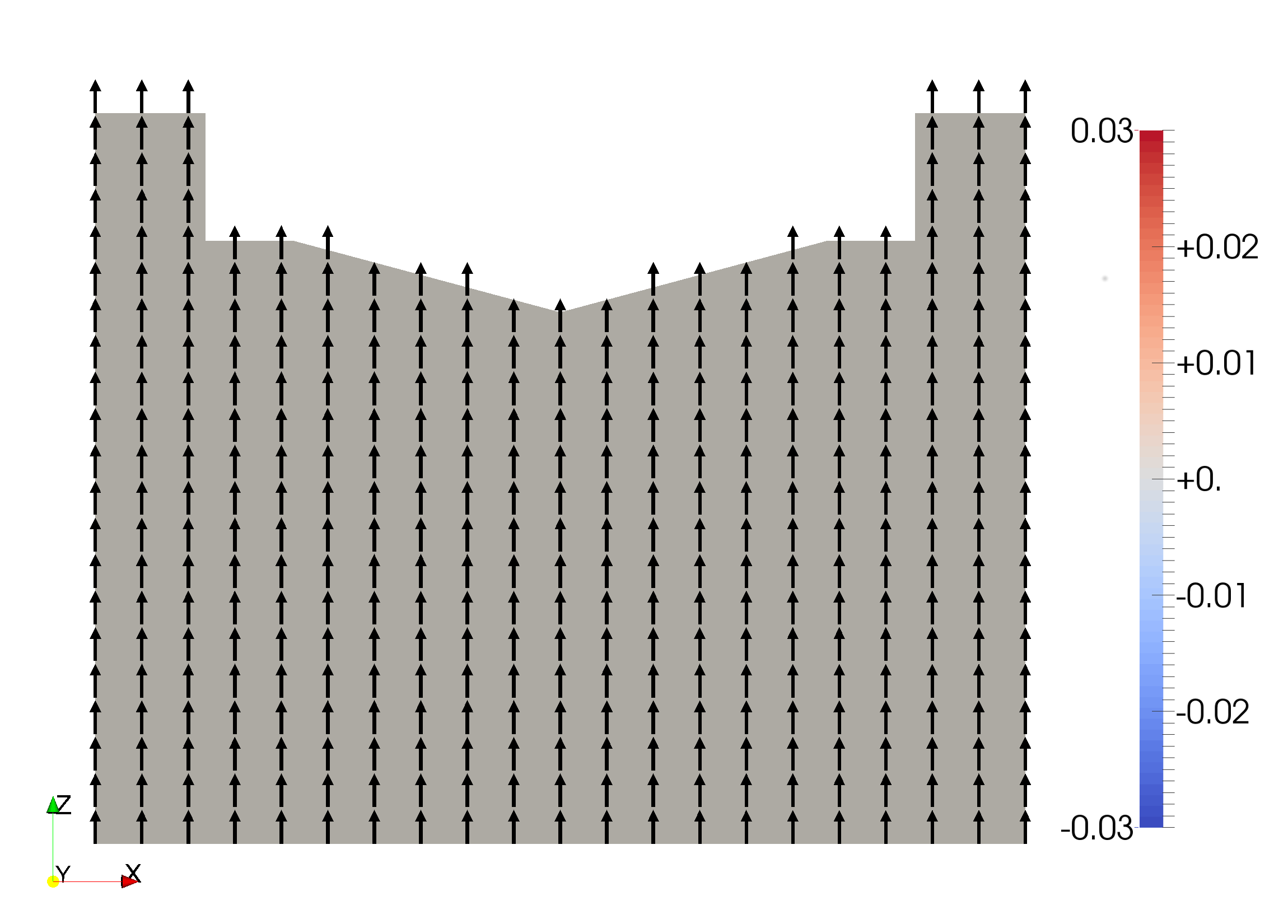}
    \put(86,64){\large $B_{rx}$}
  \end{overpic}
  \begin{overpic}[width=0.6\columnwidth]{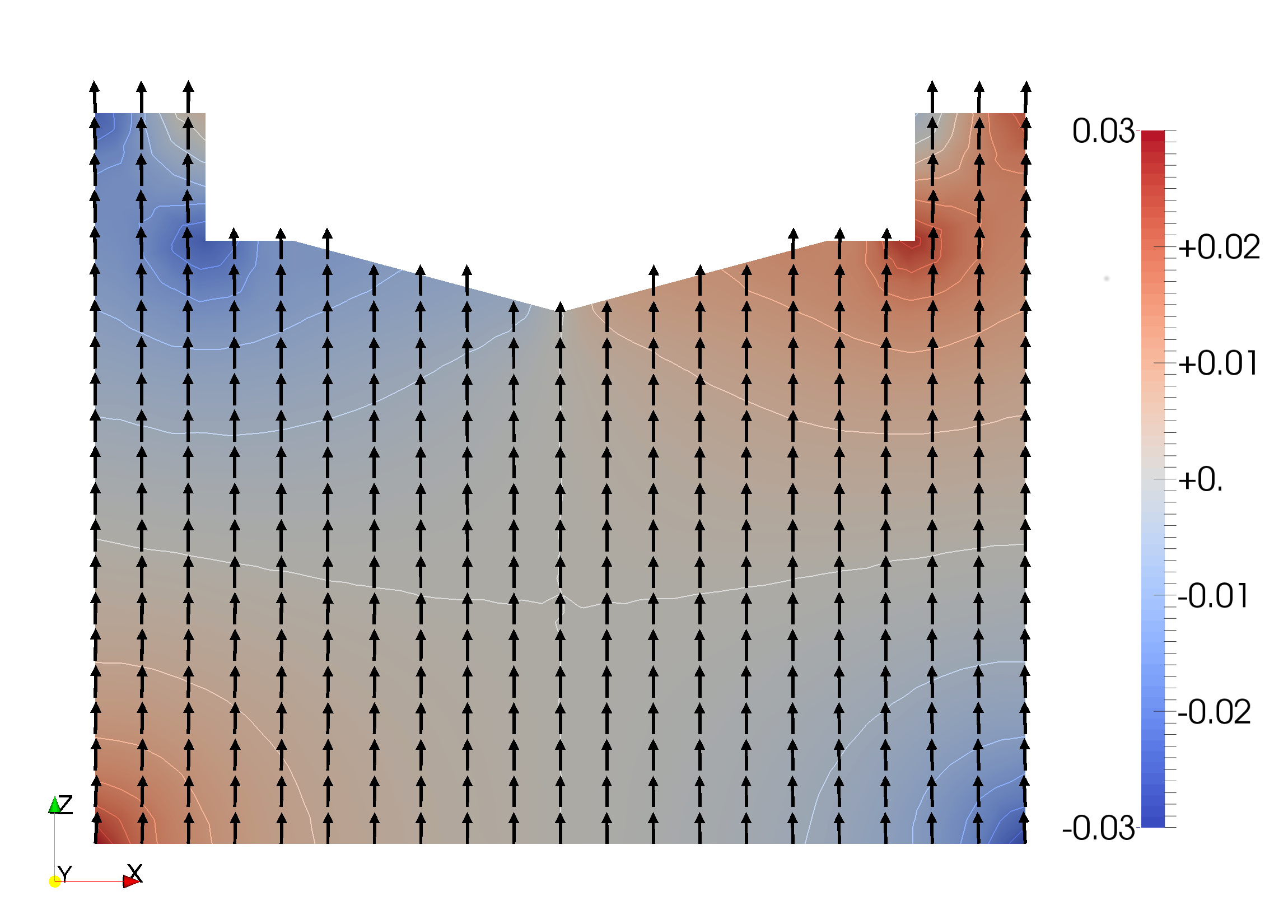}
    \put(86,64){\large $J_{x}$}
  \end{overpic}
  \caption{\small Simulation results using an anisotropic magnet model (\ref{itm:femme_anisotropic}). A homogeneous $\mathbf{B}_\text{r}$-vector is assumed within the whole magnetic region (top). The resulting magnetic polarization (bottom) shows a weak flower state, but all magnetic moments are mostly aligned with the prescribed preferred direction.}
  \label{fig:bp_anisotropic}
\end{figure}

\begin{figure}[h!]
  \centering
  \begin{overpic}[width=0.6\columnwidth]{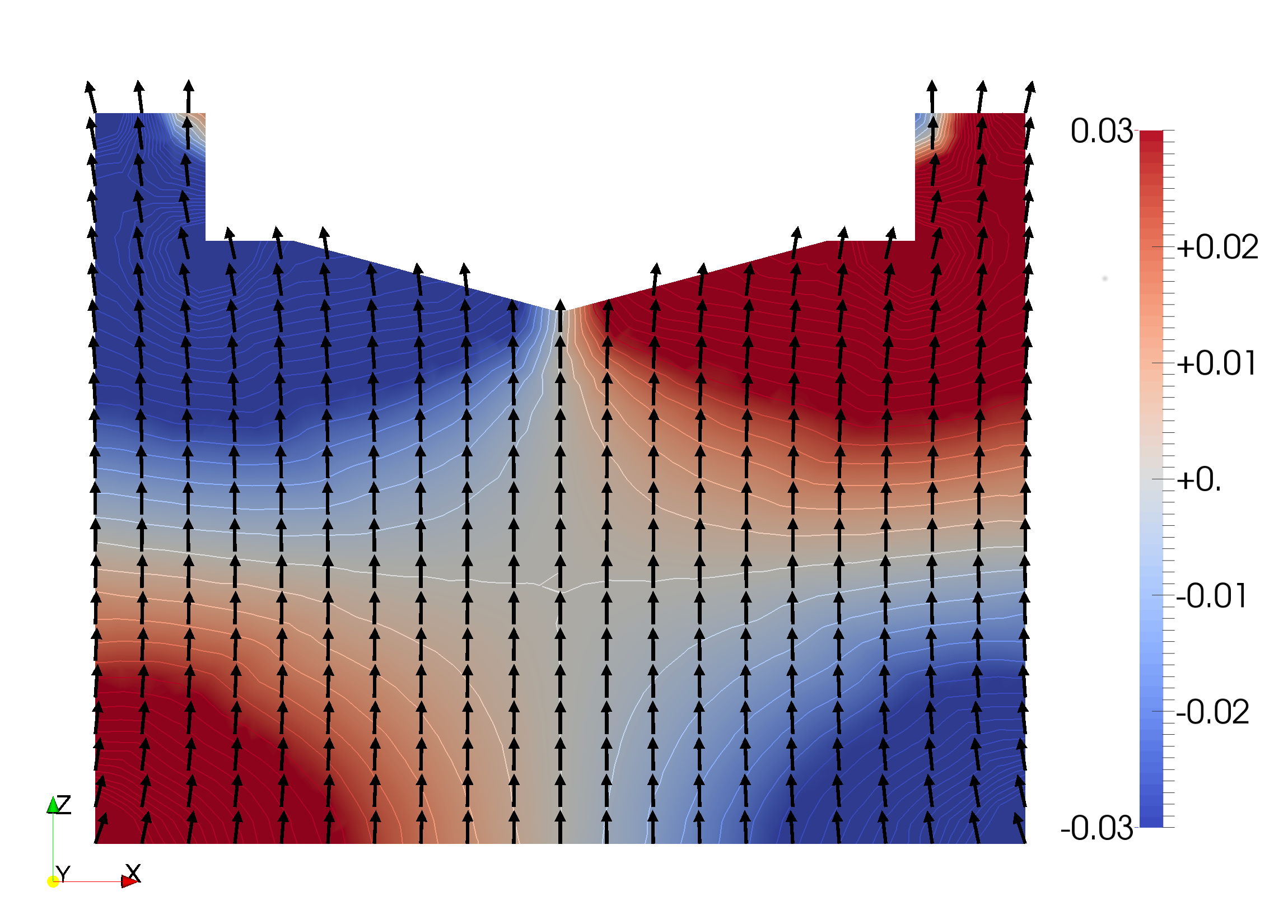}
    \put(86,64){\large $B_{rx}$}
  \end{overpic}
  \begin{overpic}[width=0.6\columnwidth]{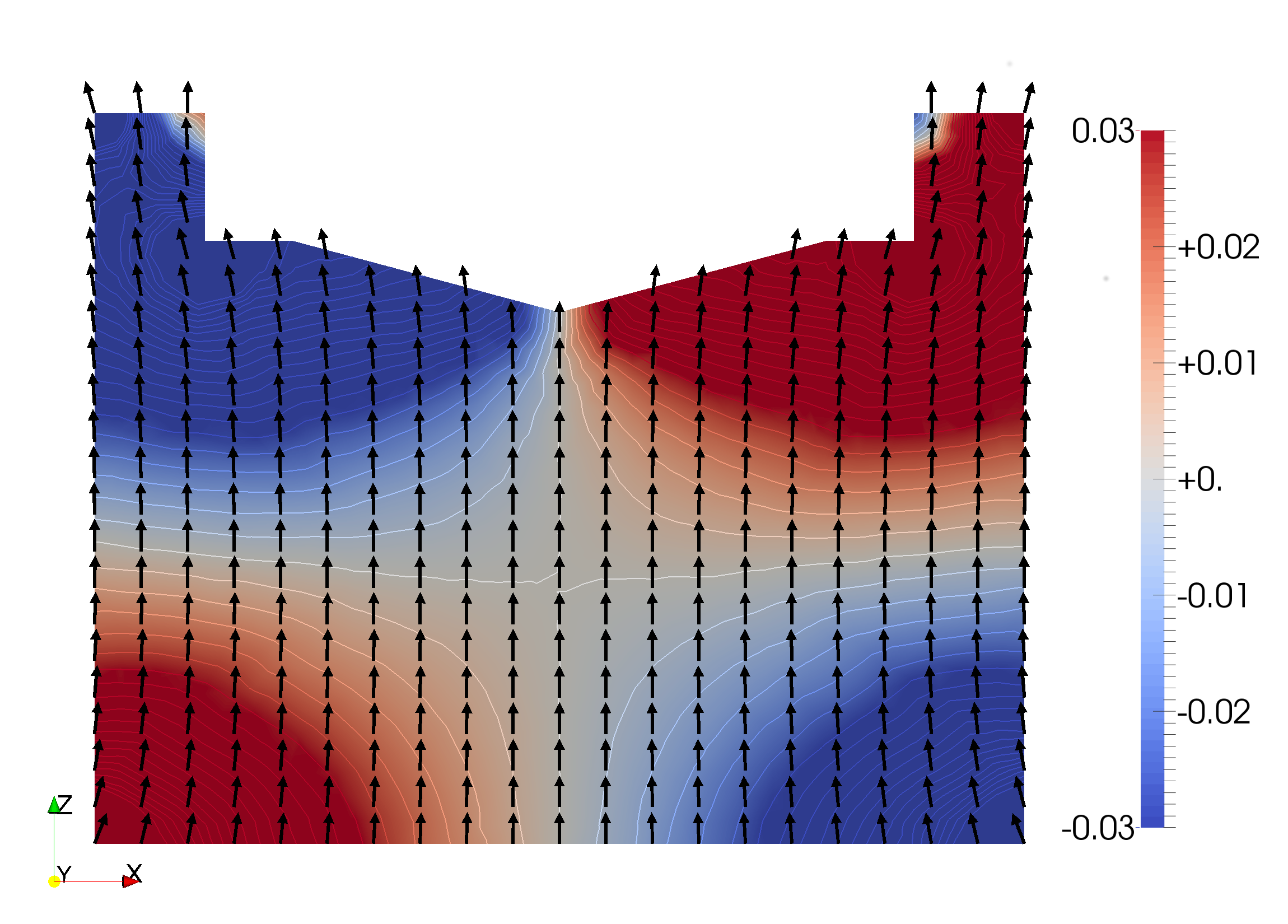}
    \put(86,64){\large $J_{x}$}
  \end{overpic}
  \caption{\small Simulation results using the presented algorithm for isotropic magnets (\ref{itm:femme_isotropic}) using a freezing field $H_\text{freeze} = 0.6$\,T. The applied field is gradually reduced from $4$\,T to $0$\,T until the final remanence state is reached. The resulting final $\mathbf{B}_\text{r}$-vector(top) shows significant deviations from the homogeneous distribution assumed for the anisotropic model. The final magnetic polarization (bottom) additionally includes the contribution from the magnetic strayfield, which further amplifies the resulting flower state.}
  \label{fig:bp_isotropic}
\end{figure}

A comparison of the anisotropic and the isotropic simulations is shown in Fig. \ref{fig:bp_anisotropic} and \ref{fig:bp_isotropic}, respectively. In contrast to the homogeneous $\mathbf{B}_\text{r}$-vector assumed for the anisotropic simulation (\ref{itm:femme_anisotropic}), the isotropic $\mathbf{B}_\text{r}$-vector state gradually degrades into a flower state while the applied field is reduced. When the external field is reduced below $H_\text{ext} \approx H_\text{freeze}$ the $\mathbf{B}_\text{r}$-vector state is completely frozen within the whole magnet. Choosing $H_\text{freeze} = 0.6$\,T yields good agreements with measurement data for various geometries. Due to the shape of the magnetic strayfield the remanence-freezing effect becomes visible mostly on sharp edges of the geometry.

Finally the comparison of the magnetic strayfield above the magnetized magnet is shown in Fig. \ref{fig:ibb_freeze}. The anisotropic models simulated with FEMME (\ref{itm:femme_anisotropic}) and ANSYS (\ref{itm:ansys_anisotropic}) coincide very well, which validates the implementation of the anisotropic model presented in Chapter \ref{ch:anisotropic_magnets} within FEMME. Furthermore it can be seen that the measurement data shows significant deviations from the anisotropic behaviour. The simulations using the isotropic magnet model (\ref{itm:femme_isotropic}) presented in Chapter \ref{ch:isotropic_magnets}, leads to a much better description of the measurement results. The derived flower state for the $\mathbf{B}_\text{r}$-vector state amplifies the influence of the magnetic strayfield onto the actual remanent magnetization.

\section{Conclusion}
An efficient method for the description of isotropic magnetic materials is presented. Comparison with measurement data reveals that in contrast to anisotropic magnets the magnetization process has a significant influence on the final remanent magnetization of isotropic magnets. This effect can approximately be described by the introduced remanence-freezing method, which leads to a very good agreement with the measurement results. The method can easily be integrated into existing anisotropic magnetic simulations. Additionally no detailed hysteresis measurements of the magnetic material are required, which simplifies the utilization for practical applications.

\section*{Acknowledgements}
The authors would like to thank the Austrian Federal Ministry of Economy, Family and Youth and the National Foundation for Research, Technology and Development for the financial support.

\bibliographystyle{ieeetr}
\bibliography{refs}
\end{document}